\documentclass[conference]{IEEEtran}
\IEEEoverridecommandlockouts
\usepackage{dblfloatfix}    

\usepackage{hyperref}
\usepackage{cite}
\usepackage{amsmath,amssymb,amsfonts}
\usepackage{algorithmic}
\usepackage{graphicx}
\usepackage{textcomp}
\usepackage{xcolor}
\def\BibTeX{{\rm B\kern-.05em{\sc i\kern-.025em b}\kern-.08em
    T\kern-.1667em\lower.7ex\hbox{E}\kern-.125emX}}
\usepackage{subcaption}
    
\usepackage[linesnumbered,ruled]{algorithm2e}

\newcommand{\I}{\text{\textnormal{I}}}

\renewcommand{\hat}{\widehat}
\renewcommand{\rho}{\varrho}

\newcommand{\E}{\mathbb E}
\renewcommand{\P}{\mathbb P}

\newcommand{\bx}{\boldsymbol{p}}
\newcommand{\bz}{\boldsymbol{z}}
\newcommand{\by}{\boldsymbol{f}}
\newcommand{\ba}{\boldsymbol{a}}

\newcommand{\bmu}{\boldsymbol{\mu}}

\newcommand{\bX}{\boldsymbol{P}}

\newcommand{\bY}{\boldsymbol{F}}

\begin{document}

\title{Ranking transmission lines by overload probability using the empirical rate function 
}

\author{\IEEEauthorblockN{Brendan Patch}
\IEEEauthorblockA{\textit{CWI} \\
Amsterdam, The Netherlands \\
brendan.patch@cwi.nl}
\and
\IEEEauthorblockN{Bert Zwart}
\IEEEauthorblockA{\textit{CWI, IEEE Member} \\
Amsterdam, The Netherlands \\
bert.zwart@cwi.nl}
}

\maketitle

\begin{abstract}
We develop a non-parametric procedure for ranking transmission lines in a power system according to the probability that they will overload due to stochastic renewable generation or demand-side load fluctuations, and compare this procedure to several benchmark approaches. Using the IEEE 39-bus test network we provide evidence that our approach, which statistically estimates the rate function for each line, is highly promising relative to alternative methods which count overload events or use incorrect parametric assumptions.
\end{abstract}


%
\section{Introduction} 
Power grid operators must determine optimal power flows while considering the probability of transmission line overload events \cite{bienstock2014chance,summers2014stochastic,dall2017chance}. 
Given power generation data corresponding to an operating point that ensures line overloads occur with very small probability, we rank lines in terms of overload probabilities.  This is useful for facilitating an efficient allocation of scarce resources to improve network resilience. 

Assume that a practitioner's objective is to rank lines according to overload probability. This is difficult for two key reasons. 
Firstly, since overload events are generally rare for all lines, systems may need to be observed for a very long time before differences between lines can be revealed using the standard method of counting the relative frequency of overload events. 
Secondly, parametric assumptions can improve the efficiency of ranking but it is not always clear which parametric assumptions are appropriate, and using the wrong parametric assumption is often highly detrimental. 
Although ranking is well studied (e.g., \cite{hall2010ranking,fan2016ranking}), results overcoming these hurdles, i.e, applicable to very small tail probabilities with an unknown distribution, are limited.

We propose a non-parametric approach to ranking lines which uses a statistical estimate of the cumulant generating or rate function (see, e.g., \cite{duffy2005large,rohwer2015convergence,duffy2017ldp} for related work) of the distribution of power line flows. 
Using the IEEE 39-bus system as given in the Matpower Simulation Package (MSP) \cite{zimmerman2010matpower}, we evaluate our proposed approach using simulated power injection data generated from Gaussian and Laplace distributions.  
For the majority of our study we model power line flows using the DC approximation (e.g., \cite{purchala2005usefulness,stott2009dc}), however we do briefly touch upon the AC model as well. 
We compare our approach with the benchmarks in terms of the probability of inaccurately identifying the set of lines which are most likely to overload (a quantity it is desired to keep small) and in terms of how accurate outputted ranks are for different numbers of observations. 

The first benchmark computes the proportion of time intervals where an overload event has occurred and then ranks the lines using this quantity as a proxy for overload probability.  
The second benchmark assumes power fluctuations are Gaussian (see e.g., \cite{bienstock2014chance,berg2016gaussian,kolumban2017short} for justification) and then uses maximum likelihood estimation (MLE) of parameter values and numerical tail probabilities to rank the lines. 
It is sensible to consider other distributions (see, e.g., \cite{mucke2011atmospheric,milan2013turbulent} for justification), hence our third benchmark assumes power fluctuations follow a Laplace distribution. 
This is highly computationally intensive when dealt with directly (e.g., using Monte Carlo (MC)), hence we establish a large deviation principle (LDP) (see, e.g., \cite{touchette2009, dz1998, bookBigQueues}) to ranking, a result which is of independent interest. 
This extends work in \cite{nesti2018emergent} where the LDP approach to ranking was shown to work very well if fluctuations follow a Gaussian distribution. 

In this paper we present two additional contributions. Firstly, our proposal to use a statistical estimate of the rate function for ranking transmission lines by overload probability is, to the author's knowledge, novel. 
Secondly, our numerical experiments illustrate key weaknesses of the benchmark methods when compared with our method. 
We see that using the proportion of intervals where an overload has occurred to estimate ranks requires a substantial amount of observations to be accurate, and for smaller numbers of observations (e.g., ${<}10^3$) does not perform as well as our method. 
For larger numbers of observations (e.g., $10^6$) we provide strong evidence that our method is still superior when more than a handful of lines need to be ranked.  
Additionally, using the incorrect parametric assumption can lead to highly inaccurate rankings, even when a large amount of data is available; an issue which our approach is not susceptible to. 
Although the majority of our analysis is performed under the DC approximation, we present some evidence that our method can still predict the top overloaded lines using a full AC model, though this should be seen as a preliminary insight.

The remainder of this paper is organized as follows. Section~\ref{sec:Model} presents our model and ranking framework. Section~\ref{sec:CS} presents a brief case study, and then Section~\ref{sec:Out} describes future research opportunities. 

\section{Model, Framework and Benchmarks}\label{sec:Model}
We model a power grid by a connected graph $(\mathcal N, \mathcal E)$ where $\mathcal N = \{1,\dots,b\}$ are nodes representing buses and $\mathcal E\subset \mathcal N \times \mathcal N$, with indices $\{1,\dots,m\}$, are directed edges representing transmission lines. Let $\mathcal S \subset \mathcal N$ be the indices of buses housing stochastic renewable power generators or unpredictable demand-side loads and $\mathcal D \subset \mathcal N$ be the indices of buses housing deterministic conventional generators. Nominal net power injections at stochastic and deterministic nodes are denoted, respectively, by $\bmu = (\mu_i)_{i\in\mathcal S}$ and $\ba=(a_i)_{i\in\mathcal D}$. 

Power injections during a time interval of interest are assumed to occur according to a random vector $\bX = (P_i)_{i\in\mathcal S}$. We assume that the distribution of $\bX$ is unknown but that we have independent and identically distributed (iid) observations 
\[
\bX_t \equiv \big(X_{1,t},\dots, X_{d,t}\big)\,,\quad t=1,\dots,n\,,
\]
each following the same distribution as $\bX$. Using the DC approximation, net line power flows and their nominal values are given by 
\begin{equation}\label{eq:DCapprox}
\bY = V_s\bX+V_d\ba\,,\quad \text{and}\quad \boldsymbol \nu = V_s\bmu + V_d\ba\,,
\end{equation}
where $V_s$ and $V_d$ are matrices encoding the grid topology and parameters. As in, for example \cite{cetinay2016topological,nesti2018emergent}), the model we use \eqref{eq:DCapprox} follows a standard setup. Choosing an arbitrary but fixed orientation of the transmission lines, the network structure displayed in Fig.~\ref{fig:ieee} is described by the edge-vertex incidence matrix $C\in \mathbb R^{m\times b}$ with entries
\[
C_{\ell,i} = \left\{\begin{array}{ll} 1 & \text{if } \ell = (i,j)\,,\\ -1 & \text{if } \ell = (j,i)\,,\\ 0 & \text{otherwise.} \end{array}\right.
\]
The parameters $\beta_\ell = \beta_{i,j} = \beta_{j,i} >0$ correspond to the susceptance of the transmission lines. By convention $\beta_{i,j} =\beta_{j,i}=0$ if there is no transmission line between $i$ and $j$. In the MSP case study these quantities are not given directly, rather they are computed from the reactance $y_\ell$ and tap ratio $\rho_\ell$ parameters of each line (which are provided in the MSP) as follows: if the tap ratio is $0$, then $\beta_\ell = y_\ell^{-1}$ and otherwise $\beta_\ell = (\rho_\ell y_\ell)^{-1}$. Store the susceptance parameters in the matrix $B = \text{diag}([\beta_1, \dots, \beta_m])$. 

Using the matrices $C$ and $B$ just defined, the network topology and weights are simultaneously encoded in the weighted Laplacian matrix of the graph $(\mathcal N, \mathcal E)$, defined as $L = C^\top B C$ or entry-wise as
\[
L_{i,j} = \left\{\begin{array}{ll} -\beta_{i,j} & \text{if } i\ne j\,,  \\ \sum_{k\ne j} \beta_{i,k} & \text{if } i = j\,. \end{array}\right. 
\]

Under the DC approximation, the relationship between any zero-sum vector of power injections $\boldsymbol x \in \mathbb R^b$ and the phase angles ${\boldsymbol \theta}\in\mathbb R^b$ they induce in the network nodes can be written in matrix form as $\boldsymbol x = L \boldsymbol \theta$. Defining $L^+ \in\mathbb R^{b\times b}$ as the Moore--Penrose pseudo-inverse of $L$ (which is easily found using most linear algebra packages), we can rewrite this as $\boldsymbol \theta = L^+ \boldsymbol x$.
This is useful in our context since it holds for any vector of power injections $\boldsymbol x \in\mathbb R^b$, even if it has non-zero sum. Since the real line power flows $\by\in\mathbb R^m$ are related with the phase angles $\boldsymbol \theta$ via the linear relation $\by = B C \boldsymbol\theta$, we therefore write $\by = BCL^+\boldsymbol x$, and hence $V = BCL^+$.
Since we consider stochastic and deterministic nodes separately, the matrices $V_s$ and $V_d$ consist of the columns of $V$ indexed by $\mathcal S$ and $\mathcal D$ respectively. 

From the power injection data $(\bX_t)_{t=1}^n$ and model \eqref{eq:DCapprox} we have line flow data $(\bY_t)_{t=1}^n$. 
A power line overloads when the absolute amount of power flowing in it exceeds a predefined threshold. Let $\gamma_1,\dots, \gamma_m>0$ denote these thresholds, and $\theta_\ell = \P(|F_\ell|\ge\gamma_\ell)\,\quad \ell\in\mathcal E$, be line overload probabilities. 
Let $\theta_{(1)} > \cdots > \theta_{(m)}$ be the ordered values of the set of parameters $\theta_1, \dots, \theta_m$, then $\theta_\ell=\theta_{(r)}$ implies that $R_\ell=r$ is the rank of line $\ell$ in this ordering. 
For $k\le m$ the lines which have the highest overload probabilities are $\mathcal R_k = (\ell\in\mathcal E : R_\ell \le k)$.
Using the data $(\bX_t)_{t=1}^n$ and different assumptions on $\bX$ there are numerous ways to generate estimates $\Theta_{\ell,n}$ of $\theta_\ell$ for $\ell \in\mathcal E$ and then based on $\Theta_{(1),n} \ge \cdots \ge \Theta_{(m),n}$ give an estimate $\mathcal R_{j,n}=\{R_{1,n},\dots,R_{j,n}\}$ with $j\ge k$ and the aim that $\mathcal R_k\subset \mathcal R_{j,n}$. 

For $\ell\in\mathcal E$, the cumulant generating function of $|F_\ell|$ is 
\[
\Lambda_{\ell}(\lambda) = \log \E\,\exp\left(\lambda |F_\ell|\right)\,,\quad\lambda\in\mathbb R\,.
\] 
Using Markov's inequality, for every $\lambda$, 
\[
\mathbb P(|F_\ell| \ge \gamma) \le \exp\left(-\lambda \gamma_\ell\right)\E\exp\left(\lambda |F_\ell|\right)\,,
\]
 so that upon letting 
\begin{equation}\label{eq:Zconconj}
J_\ell = \sup_{\lambda}\left\{\lambda\gamma_\ell-\Lambda_{\ell}(\lambda)\right\}
\end{equation}
we have $\theta_\ell \le \exp(-J_\ell)$. 

This motivates us to determine an estimate of $\mathcal R_{k}$ by directly estimating the rate function of $|F_\ell|$, for each $\ell\in\mathcal E$, from the data $(\bY_t)_{t=1}^n$. That is, for $\lambda \in \mathbb R$ we compute $\hat\Lambda_{\ell}(\lambda) = \log \frac{1}{n}\sum_{t=1}^n \exp\left(\lambda |F_{\ell,t}|\right)$, with corresponding estimate of \eqref{eq:Zconconj}, 
\begin{equation}\label{eq:ZconconjEst}
\hat J_{\ell,n} = \max_{\lambda \in \mathbb R}\left\{\lambda\gamma_\ell-\log \frac{1}{n}\sum_{t=1}^n \exp\left(\lambda |F_{\ell,t}|\right)\right\}\,,
\end{equation}
and then use $\Theta_{\ell,n} = \exp\left(-\hat J_{\ell,n}\right)$ for $\ell\in\mathcal E$ to estimate $\mathcal R_{j,n}$. For large $n$ the solution to \eqref{eq:ZconconjEst} will not change much when new data is added, allowing it to be solved online quickly with standard methods. We call this approach Alg.~1.

In order to benchmark our approach, we now describe three other ways to estimate $\mathcal R_{k}$ using the data $(\bX^{(t)})_{t=1}^n$ and model \eqref{eq:DCapprox}. 

{\em Alg. 2: Empirical distribution function.} Firstly, we can work with $\I\{|F_\ell|\ge\gamma_\ell\}$, which evaluates to 1 when $\{|F_\ell|\ge\gamma_\ell\}$ occurs and is $0$ otherwise, and estimate $\theta_\ell$ using 
\[
\Theta^{(n)}_{\ell} = \frac{1}{n}\sum_{t=1}^n \I\{|Y_{\ell,t}|\ge\gamma_\ell\}\,.
\] 
The collection $(\Theta_{\ell,n})_{\ell=1}^m$ can then be used to estimate $\mathcal R_{j,n}$. This approach is straightforward to implement and does not rely on any potentially erroneous assumptions.  
However, since it relies on observations of rare overload events in order to make predictions and is therefore unlikely to perform well when data on such events is limited. 


{\em Alg.~3: Gaussian assumption benchmark.} Another approach is to assume that $\bX$ follows a Gaussian distribution with known mean $\bmu$ and covariance $\Sigma$.  
Since the space of multivariate Gaussian distributions is closed under affine transformations, it is easily seen that under this assumption $F_\ell$ follows a normal distribution with mean $\nu_\ell$ (as given in \eqref{eq:DCapprox}) and variance given by $(V_s\Sigma V_s^\top)_{\ell,\ell}$. 
Using $\hat \Sigma = \frac{1}{n}\sum_{t=1}^n (\bX_t-\bmu)(\bX_t-\bmu)^\top$, the MLE of $\Sigma$, it is then straightforward to numerically determine estimates of $\theta_\ell$ for each $\ell$ and a corresponding estimated ranking $\mathcal R_{j,n}$. 


{\em Alg.~4: Laplace assumption benchmark.} It is useful to consider an alternative to the Gaussian distribution since this allows us to investigate what can go wrong with parametric estimation procedures when the incorrect distribution is used. 
We may, for example, assume for each $i\in\mathcal V$ that $P_i$ follows a Laplace distribution with known mean $\mu_i$ but unknown scale $\varepsilon\alpha_i$, specifically, that $P_i$ has probability density function $h_{i}(x) = (2\varepsilon\alpha_i)^{-1}\exp\left(-(\varepsilon\alpha_i)^{-1}|x-\mu_i|\right)$, for $x\in\mathbb R$. 

This distribution corresponds to the difference of two iid exponential distributions with expected difference $\mu_i$. 
The variable $\varepsilon$ is included here since we suppose that $(\alpha_i)_{i\in\mathcal V}$ must be estimated from data and consider a small noise approximation where $\varepsilon\to 0$. 
We use this small noise approximation since determining $\theta_\ell$ numerically or using simulation is highly computationally intensive --- making such approaches inappropriate for use in a large number of benchmarking experiments. 
The MLE of $\alpha_i$ is given by the average of the absolute deviations from the expected value, $\hat \alpha_i = \frac{1}{n}\sum_{t=1}^n \big|P_{i,t}-\mu_i\big|$.
This estimator is more robust to outliers in the data than the Gaussian distribution approach (as the tail is heavier). 
A sequence of random vectors $(\bX^{(\varepsilon)})_{\varepsilon\in\mathbb R}$ taking values in $\mathbb R^d$ is said to satisfy a large deviations principle (LDP) with rate function $H : \mathbb R^d \to \mathbb R$ if $H$ satisfies some minor technical conditions (see, e.g., \cite[p.~4--5]{dz1998}) and if for any measurable set $\Gamma\subset \mathbb R^d$ it roughly holds that 
\begin{equation}\notag
\lim_{\varepsilon\to0}\varepsilon\log \P\big(\bX^{(\varepsilon)}\in\Gamma\big) = -\min_{\bx\in\Gamma}~H(\bx)\,. 
\end{equation}
Under the assumption that $\bX$ is Laplacian as described above, this random vector satisfies an LDP with rate function $H(\bx) = \sum_{i\in\mathcal V} \alpha_i^{-1}|p_i-\mu_i|$. The proof of this result is standard --- for half spaces the rate function follows immediately from the form of the cumulative distribution function, and for more general sets the result follows similar arguments to the proof of Cram\'er's theorem (see, e.g, \cite[pp.~33--36]{bookBigQueues}). 

Using this LDP, we can estimate $\theta_\ell$ by $\Theta_{\ell,n} = \exp\left(-H_{\ell,n}/\varepsilon\right)$, where $H_{\ell,n}$ is the optimal value of the linear programming problem 
\begin{equation}\label{eq:OPT2}
\begin{array}{cl} \min\limits_{\bx,\bz\in\mathbb R^d} &\bz^\top\hat{\boldsymbol \alpha}^{-1}\,, \\
\text{s.t.} &  \text{sign}(\nu_\ell)(V_{s,\ell}\bx+V_{d,\ell}\ba) \ge \gamma_\ell\,,\\
& z_i \ge \alpha_i^{-1}(p_i-\mu_i)\,,\quad i \in \mathcal V\,, \\
&  z_i \ge -\alpha_i^{-1}(p_i-\mu_i)\,,\quad i \in \mathcal V \,,
\end{array}
\end{equation}
where $V_{s,\ell}$ and $V_{d,\ell}$ are the $\ell$-th row of $V_s$ and $V_d$ respectively, and where $\varepsilon$ is chosen small (note that since $\varepsilon$ is equal for all lines, that the value of this parameter does not affect the ranking). This program arises using a standard technique for optimization problems with absolute values in their objective function (see, e.g., \cite[Ch.~6]{boyd2004convex}). 
The sign function arises due to symmetry; which holds since sums of independent symmetric random variables are necessarily symmetric. 
This collection can then be used to determine an estimate $\mathcal R_{j,n}$. Although we treat this algorithm as a benchmark, it appears to be new. 


\section{Case study: IEEE 39 Network}\label{sec:CS}
\begin{figure}
\centering 
\includegraphics[width=\linewidth]{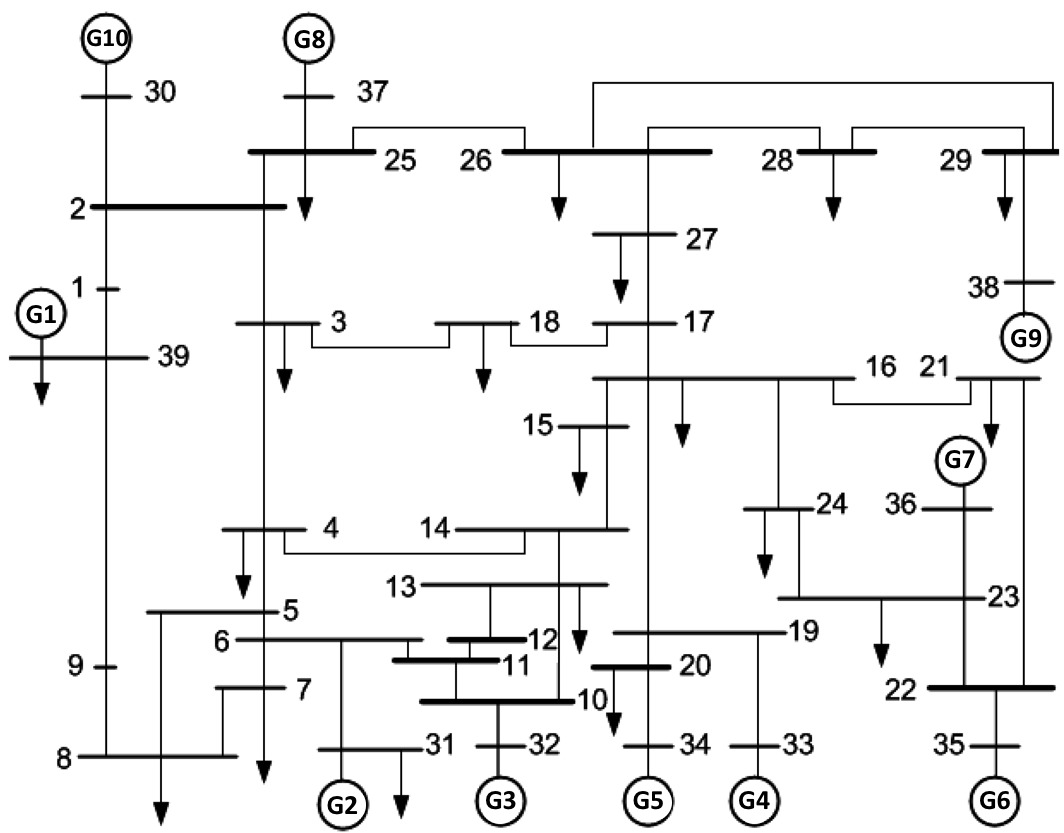}
\caption{IEEE 39-bus New England interconnection system.}
\label{fig:ieee}
 \end{figure}

Consider the IEEE 39-bus New England interconnection system, displayed in Fig.~\ref{fig:ieee}, which has $10$ generators and $29$ load nodes connected by a network of $46$ lines with susceptance parameters as given in the MSP. We utilize the Python 3.7.5 SciPy minimize function using the `SLSQP' routine to solve the optimization problem in \eqref{eq:ZconconjEst} (which in this unconstrained setting and considering we do not explicitly provide a derivative reduces to the secant method). We utilize the Python 3.7.5 SciPy linprog function using both the `interior-point' and `revised simplex' routines to solve \eqref{eq:OPT2}, and then if there is a difference in solution we use the one with the smaller objective value. This case study is available online at \href{https://github.com/bpatch/power-system-line-rank}{https://github.com/bpatch/power-system-line-rank}. 

We test our method using simulated power flow injection data generated according to either a Gaussian or a Laplace distribution, as detailed in Section~\ref{sec:Model}. We assume that all 10 generators are stochastic and that all other nodes are deterministic. Although we use realistic parameter values, the specific details of the stochastic model in this preliminary case study are illustrative only; in our follow up work we will use real data (as in, e.g., \cite{nesti2018emergent}).  

In the Gaussian case, each generator has nominal net power injection $\mu_i$ given by nominal power generation minus nominal demand as given in the MSP (where the optimal power flow has already been solved). The variance of net power injection is $5$ times the value of nominal injections. For example, generator $2$ has a nominal injection of $677.87$MW and nominal demand of $9.2$MW so it has $\mu_i = 668.67$ and Var$(P_i) = 3389.35$.
The off diagonal entries of $\Sigma$ are extracted from $AA^\top$ where $(A)_{i,j}$ are generated iid from a mean zero Gaussian distribution with variance $25$.  In \cite{nesti2018emergent} evidence is given, based on a dataset described in \cite{horsch2018pypsa} for the German power grid, that the mean standard deviation of solar power generation is approximately $9\%$ of installed capacity; in our small case study here we remain very close to this finding with a mean standard deviation of stochastic power generation of $9.3\%$ of installed capacity. Given this parameterization of $\bX$, we determined that line 27 is most likely to overload with $\theta_{27}\approx 0.017622$, followed by line 3 with $\theta_3 \approx  6{\cdot}10^{-6}$.

For the simulated data generated according to a Laplace distribution we used the same values of $\mu_i$ and Var$(P_i)$ as for the Gaussian case. Since the variance of a Laplace distribution with scale $\alpha_i$ is given by $2\alpha_i^2$, we have $\alpha_i = \sqrt{\text{Var}(P_i)/2}$. In this case, from extensive MC simulations as well as Alg.~4 with perfect knowledge we again have that line 27 is most likely to overload, now with $\theta_{27} \approx 0.023863$. Interestingly, the line which is second most likely to fail is now line 20 with $\theta_{20}\approx 0.000841$, indeed $R_3=7$ according to Alg.~4 with perfect information (and $R_3=6$ according to a MC estimate with $10^{10}$ samples).  This is evidence that incorrect distributional assumptions can have a substantial impact on the true ranking of lines according to overload probabilities. 

{\em Experiment 1.} Our first focus is on determining the accuracy of the estimate $\mathcal R_{j,n}$. In particular, we investigate the probability of false selection $f^{n}_{k,j} = \P(\mathcal R_k \not\subset \mathcal R_{j,n})$ when $\mathcal R_{j,n}$ is generated using our proposed approach and benchmarks.

\begin{figure}[b!]
\centering 
\begin{subfigure}[b]{\linewidth}
	\includegraphics{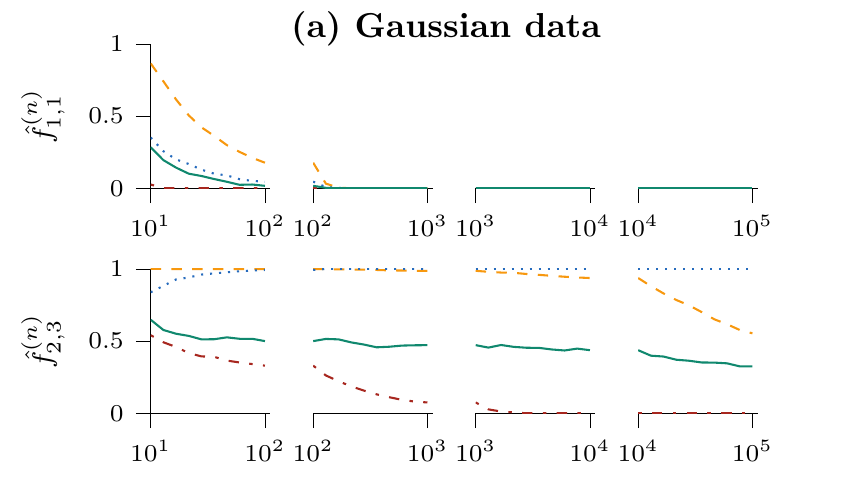}
\end{subfigure}
\begin{subfigure}[b]{\linewidth}
	\includegraphics{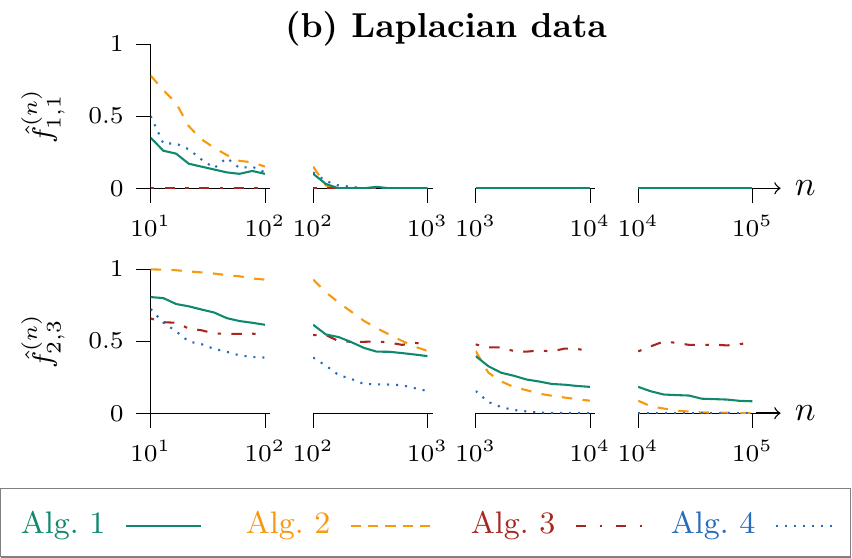}
\end{subfigure}
\caption{Estimated probability of false selection for our approach (Alg.~1) compared with benchmarks, as a function of sample size when $\bX$ is: (a) Gaussian distributed, or (b) Laplace distributed. (Note: linear scale within each plot.) }
\label{fig:PFS}
 \end{figure}

Fig.~\ref{fig:PFS}a plots estimates of the probability of false selection for $k=j=1$ (top panel) and for $k=2$ and $j=3$ (bottom panel), as a function of the number of observations $n$ in the simulated Gaussian data $(\bX^{(t)})_{t=1}^n$. The estimates $\hat f_{1,1}^{n}$ and $\hat f_{2,3}^{n}$ of $f_{1,1}^{n}$ and $f_{2,3}^{n}$ for our algorithm and benchmarks are produced for a range of $n$ by determining the proportion of samples out of $10^3$ that correctly estimate $\mathcal R_{1,n} = \mathcal R_{1} = \{27\}$ and, respectively, $\mathcal R_{3,n} \supset \mathcal R_2 = \{27,3\}$. For each sample the same dataset is used by all four algorithms to produce these estimates, the dataset for $n$ is the dataset for $n-1$ with an additional observation added. 

The plot of $\hat f_{1,1}^{n}$ in Fig.~\ref{fig:PFS}a shows that when it is known that the data is Gaussian only 10 observations are needed to achieve near complete accuracy of the estimate of $\mathcal R_1$, while if the data is (wrongly) assumed to be Laplace distributed, then after approximately $10^2$ observations $\hat f_{1,1}^n$ is close to $0$. Our approach, which makes no distributional assumption is more accurate than when the wrong assumption is made, but also only approaches $0$ after approximately $10^2$ observations. The approach based on counting failure events performs less well, requiring approximately 200 samples to achieve a very high level of accuracy. The plot provides evidence that making the correct parametric assumption greatly assists the estimation procedure. Indeed, in this case making the wrong parametric assumption still appears to be better than not making any parametric assumption at all. We highlight that our proposed empirical rate function based method performs vastly better than the naive counting approach.  

The plot of $\hat f_{2,3}^{n}$ in Fig.~\ref{fig:PFS}a provides evidence that in this example $\mathcal R_2$ can be much more difficult to identify than $\mathcal R_1$. Alg.~3, which makes the correct distributional assumption, is accurate once $n$ reaches approximately $10^3$. Of more significance, however, is the fact that wrongly assuming the data follows a Laplace distribution result in a catastrophic failure of the ranking procedure (i.e., $\hat f_{2,3}^{n}$ is always close to 1 for Alg.~4). In addition, even at $n=10^5$ Alg.~2 is wrong in more than half of the instances. Alg.~1, while also never reaching high levels of accuracy, performs much better than its naive non-parametric competitor Alg.~2, especially when fewer observations are used. In summary, the plot provides evidence that making the wrong parametric assumption can result in complete failure of estimation and that our proposed empirical rate function based method Alg.~1 performs better than Alg.~2.  

Fig.~\ref{fig:PFS}b replicates the experiments presented in Fig.~\ref{fig:PFS}a, this time using the Laplace simulated data previously described. The figure provides further evidence supporting our previous observations, extending them to the Laplace case. Importantly, there is evidence that incorrect parametric assumption can lead to very poor estimates. Additionally, our empirical rate function based method Alg.~1 performs much better than the indicator function based Alg.~2 when $n$ is less than $10^3$, however in this case it then performs comparatively less well when more observations are available --- as we will show in the next experiment, this superiority is limited to cases when only a small number of rankings are required. 

\begin{figure}[b!]
\centering 
\begin{subfigure}[b]{\linewidth}
	\includegraphics{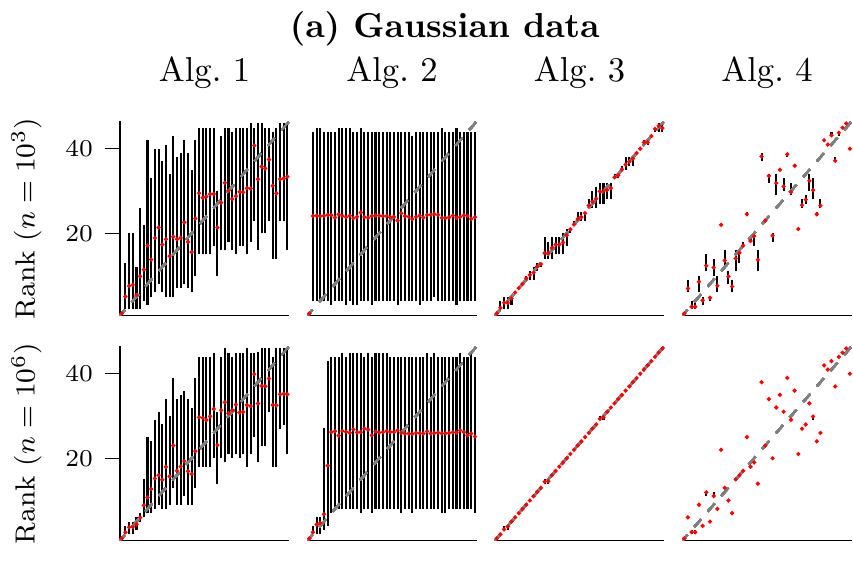}
\end{subfigure}
\begin{subfigure}[b]{\linewidth}
	\includegraphics{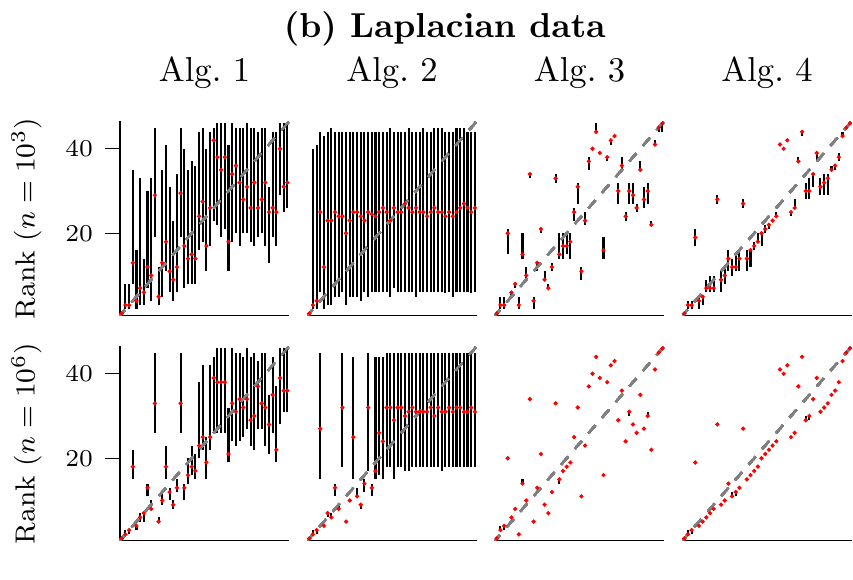}
\end{subfigure}
\caption{Simulated prediction intervals (black bars) and mean rankings (red dots) with lines are arranged by true rank (increasing left to right). }
\label{fig:Rank}
 \end{figure}

{\em Experiment 2.} Fig.~\ref{fig:Rank} displays simulated prediction intervals (black bars) and means for rankings obtained from our approach (Alg.~1) compared with benchmarks, when lines are arranged by true rank (increasing left to right) for $n=10^3$ (top rows) and $n=10^5$ (bottom rows) when $\bX$ is: (a) Gaussian distributed, or (b) Laplace distributed. A first observation is that using the incorrect parametric assumption regularly results in a low variance but highly biased estimation of rank, this is evidenced by very small prediction intervals that regularly do not contain the true rank. Indeed, for Laplace distributed data the rankings even using the algorithm which knows the data is Laplace distributed results in biased rankings.  A second observation is that of the two non-parametric procedures, our procedure performs substantially better throughout the entire range of lines, despite being weaker for identifying the top 2 lines when there is large amount of data (as observed in Exp.~1). Our procedure generally has much smaller prediction intervals that regularly contain the true rank and exhibits a much more evident positive linear relationship between true rank and mean estimated rank. Finally, in all cases the line which is most likely to overload is overwhelmingly accurately assigned rank $1$, but lines with higher rankings are substantially less reliably ranked.

{\em Experiment 3.} In this experiment we briefly explore the effectiveness of our algorithm when an AC power flow model is used in place of \eqref{eq:DCapprox}. 
Specifically, we use the network.pf() function in the PyPSA Python software toolbox \cite{brown6pypsa} to map power injection realizations to line flow realizations.  
We continue using the IEEE 39-bus test case and retain the same nominal injections $\bmu$, as well as the $\Sigma$ and $\alpha$ values from the earlier experiments. 
We directly use the reactance and resistance parameters as given in the MSP case study in the network.pf() function, rather than the susceptance parameters as used in our earlier experiments. Alg.~3 does not work in this setting, as the associated mapping is highly nonlinear. An extension of Alg.~4 to this case would constitute an application LD techniques that is beyond the scope of this paper. 

\begin{figure}[ht]
\centering 
\includegraphics{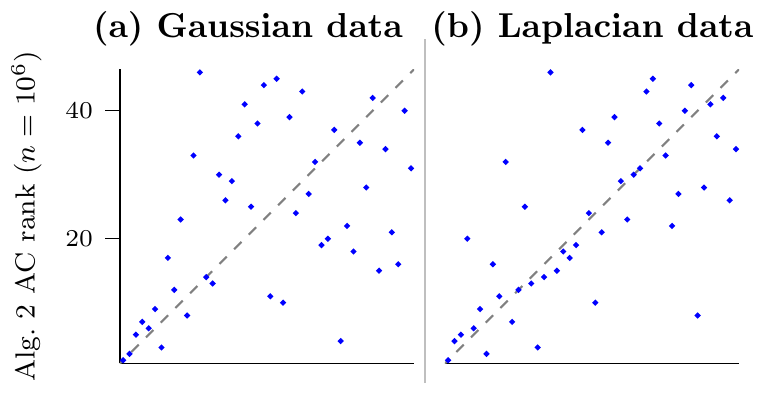}
\caption{Simulated mean rankings (blue dots) from $10^6$ observations when an AC power flow model is used in place of \eqref{eq:DCapprox} with lines arranged according to the true rank (increasing left to right) using the DC power flow model \eqref{eq:DCapprox} as outputted by: (a) Alg.~3 with known $\Sigma$, and (b) Alg.~4 with known $\alpha$.}
\label{fig:RankAC2}
 \end{figure}

Using $10^6$ observations we determined that in both the Gaussian and Laplacian cases line 27 is still the most likely line to overload; now with $\theta_{27}\approx\Theta_{27,10^6} = 0.027271$ in the Gaussian case and  $\theta_{27}\approx\Theta_{27,10^6} = 0.0328126$, both with score intervals of size less than $7.7\cdot 10^{-6}$. 
Interestingly, line 3 is now estimated to be the second most likely line to overload for both disturbance distributions. 
To investigate the difference in rankings between the AC and DC models for a broader range of ranks, in Fig.~\ref{fig:RankAC2} we arrange the lines from left to right in terms of their true rank (increasing from 1 to 46) as given by Alg.~3 and Alg.~4 with perfect knowledge of $\Sigma$ and $\alpha$ and then plot the corresponding rank given by Alg.~2 after $10^6$ observations. It can be seen for both disturbance distributions that for lines with a ranking below approximately 10 the DC and AC ranking has a strong linear relationship, suggesting a strong match between DC and AC rankings. Above rank 10 the relationship is quite weak, however, as we saw in the previous experiment, for higher rankings and $10^6$ observations Alg.~2 is not very accurate --- so for this range of rankings it is not clear if the lack of a relationship is due to noise or because a relationship does not exist. In summary, the figure provides some evidence that ranking lines based on the DC approximation can be informative about the ranking under an AC model. More research is needed to investigate this issue. 

\begin{figure}[b!]
\centering 
\includegraphics{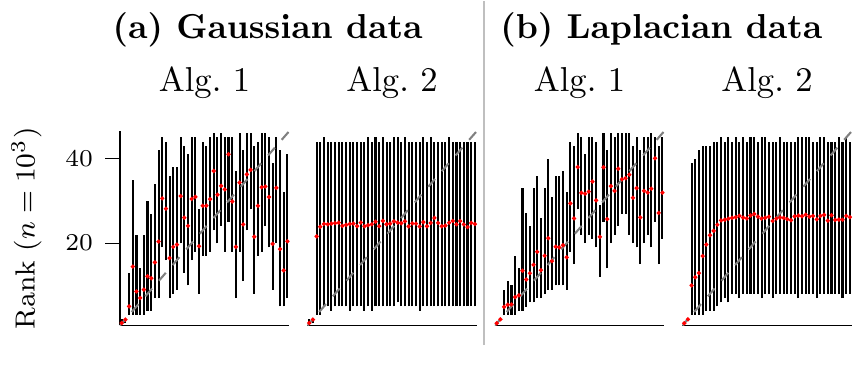}
\caption{Simulated prediction intervals (black bars) and mean rankings (red dots) with lines arranged according to the rank (increasing left to right) outputted by Alg.~2 with $10^6$ observations when an AC power flow model is used.
}
\label{fig:RankAC1}
 \end{figure}

{\em Experiment 4.} For our final experiment we see whether our approach Alg.~1 outperforms Alg.~2 when an AC model is used in place of \eqref{eq:DCapprox}. In Fig.~\ref{fig:RankAC1} lines are arranged from left to right in terms of their rank (increasing) as given by Alg.~2 after $10^6$ observations and using the AC model. We then plot mean rankings and prediction intervals based on $10^3$ MC samples of rank using Alg.~1 and Alg.~2, each based on $10^3$ observations. It can be seen that for both disturbance distributions Alg.~1 with $10^3$ observations is able to more reliably recover the Alg.~2 rank based on $10^6$ observations than Alg.~2 is with $10^3$ observations (as evidenced by smaller prediction intervals and a stronger linear relationship). 
We therefore have some evidence that Alg.~1 is effective in the AC setting and outperforms Alg.~2.  

\section{Outlook}\label{sec:Out}
We have presented a non-parametric approach for ranking power lines in terms of overload probability by statistically estimating the rate function, which in our case study appears highly promising.  We compared this with benchmarks which typically performed less well due to inefficiency in capturing data structure or relying on incorrect parametric assumptions.

There remain open questions and further research is required. Most pertinent is further exploration of our methods using more sophisticated models than \eqref{eq:DCapprox}, in particular the consideration of AC power flows. Another potentially useful avenue is the adaptation of our approach to the guidance of computing budget allocation (as in, e.g., \cite{glynn2004large,liu2019optimal}). LD theory can be applied to nonlinear mappings as well, and as such is natural to explore (e.g., to extend Alg.~4). 

\section*{Acknowledgments}
{\footnotesize This work was supported by NWO grant 639.033.413. We thank Tommaso Nesti, Andrew Richards, and Alessandro Zocca for useful discussions. }


\end{document}